%Paper: hep-ex/9410009
%From: Kevin Pitts <KPITTS@FNALI.FNAL.GOV>
%Date: Mon, 10 Oct 1994 22:22:00 -0500 (CDT)

%macropackage=phyzzx
%\input phyzzx
\input tables

\def\act{|cos\theta| }
\def\ft{f_t(\act)}
\def\gee{$\Gamma_{ee}$}

\def\Pe{\cal P\rm_e}

\def\alr{$A_{LR}$}
\def\alrbb{$A_{LR}^{\epo}(\act)$}
\def\alrraw{$\widetilde A_{LR}^{e^+e^-}(\act) $}

\def\bha{$e^+e^- \rightarrow e^+e^-$}

\def\x0{X$_0$}

\def\ep{$e^+e^-\;$}
\def\epo{e^+e^-}

\def\z0{$Z^0$}

\def\alp{\relax\ifmmode \alpha_S\else $\alpha_S$\fi$\;$}

\def\alr{A_{LR}}

\def\lum{{\cal L}}

\def\oversim#1#2{\lower0.7ex\vbox{\baselineskip 0pt plus0pt minus0pt
\lineskip 0pt plus0pt minus0pt \lineskiplimit 0pt
  \ialign{$\mathsurround=0pt #1\hfil##\hfil$\crcr#2\crcr\sim\crcr}}}
\PHYSREV
\sequentialequations
% ***********************************
% * These are for the preprint only *
% ***********************************

\vskip .2 in
\pubnum{6605}
\date{September 1994}
\pubtype={(E/T)}
\titlepage
\tolerance=10000

\vskip -.3 in
\title{Polarized Bhabha Scattering and a Precision Measurement
of the Electron Neutral Current Couplings\raise .4 ex\hbox{$\ast$}}

\vskip -.1 in
\centerline{The SLD Collaboration\raise .4 ex\hbox{$\dagger$}}

\vskip -.1 in
\centerline{\it Stanford Linear Accelerator Center}
%\medskip

\vskip -.1 in
\centerline{\it Stanford University, Stanford, California 94309}

\vskip -.2 in
\abstract{The cross section for Bhabha scattering (\bha) with
polarized electrons at the center of mass energy
of the \z0 resonance has been measured with the
SLD experiment at the SLAC Linear Collider (SLC)
during the 1992 and 1993 runs.
The first measurement of the left-right asymmetry in
Bhabha scattering ($A_{LR}^{\epo}(\act)$) is presented.
{}From               $A_{LR}^{\epo}(\act)$
the effective
weak mixing angle is measured to be
$sin^2\theta_W^{\rm eff} = 0.2245 \pm 0.0049 \pm 0.0010 $.
The effective electron
vector and axial vector couplings to the \z0 are extracted
from a combined analysis of
the polarized Bhabha scattering data   and
and the left-right asymmetry ($A_{LR}$) previously published by this
collaboration.
{}From the combined 1992 and 1993 data
the effective electron  couplings
are measured to be $v_e = -0.0414 \pm 0.0020$ and
$a_e = -0.4977 \pm 0.0045$.
% These results are compared with other experiments.
}

\vskip -.3 in
 \submit{Physical Review Letters}

\vskip -.0 in
\noindent
%$\ast$ Work supported by the Department of Energy,
%  the National Science Foundation,
%  the UK Science and Engineering Research Council,
%  the Istituto Nazionale di Fisica Nucleare of Italy,
%  and the Japan-US Cooperative Research Project on High Energy Physics.
\vbox{\eightrm\singlespace\noindent
$\ast$ Work supported by the Department of Energy
  contracts:
  DE-FG02-91ER40676 (BU),
  DE-FG03-92ER40701 (CIT),
  DE-FG03-91ER40618 (UCSB),
  DE-FG03-92ER40689 (UCSC),
  DE-FG03-93ER40788 (CSU),
  DE-FG02-91ER40672 (Colorado),
  DE-FG02-91ER40677 (Illinois),
  DE-AC03-76SF00098 (LBL),
  DE-FG02-92ER40715 (Massachusetts),
  DE-AC02-76ER03069 (MIT),
  DE-FG06-85ER40224 (Oregon),
  DE-AC03-76SF00515 (SLAC),
  DE-FG05-91ER40627 (Tennessee),
  DE-AC02-76ER00881 (Wisconsin),
  DE-FG02-92ER40704 (Yale);
  National Science Foundation grants:
  PHY-91-13428 (UCSC),
  PHY-89-21320 (Columbia),
  PHY-92-04239 (Cincinnati),
  PHY-88-17930 (Rutgers),
  PHY-88-19316 (Vanderbilt),
  PHY-92-03212 (Washington);
  the UK Science and Engineering Research Council
  (Brunel and RAL);
  the Istituto Nazionale di Fisica Nucleare of Italy
  (Bologna, Ferrara, Frascati, Pisa, Padova, Perugia);
  and the Japan-US Cooperative Research Project on High Energy Physics
  (Nagoya, Tohoku).}

\endpage

\vfill\eject
\bigskip
\centerline{{$\dagger$}\ The SLD Collaboration}
\medskip
  \def\iADEL{$^{(1)}$}
  \def\iBOL{$^{(2)}$}
  \def\iBU{$^{(3)}$}
  \def\iBRUN{$^{(4)}$}
  \def\iCIT{$^{(5)}$}
  \def\iUCSB{$^{(6)}$}
  \def\iUCSC{$^{(7)}$}
  \def\iCIN{$^{(8)}$}
  \def\iCSU{$^{(9)}$}
  \def\iCOLO{$^{(10)}$}
  \def\iCOL{$^{(11)}$}
  \def\iFER{$^{(12)}$}
  \def\iFRA{$^{(13)}$}
  \def\iILL{$^{(14)}$}
  \def\iLBL{$^{(15)}$}
  \def\iMIT{$^{(16)}$}
  \def\iMASS{$^{(17)}$}
  \def\iMISS{$^{(18)}$}
  \def\iNAG{$^{(19)}$}
  \def\iOREG{$^{(20)}$}
  \def\iPAD{$^{(21)}$}
  \def\iPERU{$^{(22)}$}
  \def\iPISA{$^{(23)}$}
  \def\iRUT{$^{(24)}$}
  \def\iRAL{$^{(25)}$}
  \def\iSLAC{$^{(26)}$}
  \def\iTENN{$^{(27)}$}
  \def\iTOH{$^{(28)}$}
  \def\iVAND{$^{(29)}$}
  \def\iWASH{$^{(30)}$}
  \def\iWISC{$^{(31)}$}
  \def\iYALE{$^{(32)}$}
  \author{                         % author and institution list
  \baselineskip=.75\baselineskip   % shrink the interline spacing
  \hbox{K. Abe                 \unskip,\iTOH}
  \hbox{I. Abt                 \unskip,\iILL}
  \hbox{T. Akagi               \unskip,\iSLAC}
  \hbox{W.W. Ash               \unskip,\iSLAC}
  \hbox{D. Aston               \unskip,\iSLAC}
  \hbox{N. Bacchetta           \unskip,\iPAD}
  \hbox{K.G. Baird             \unskip,\iRUT}
  \hbox{C. Baltay              \unskip,\iYALE}
  \hbox{H.R. Band              \unskip,\iWISC}
  \hbox{M.B. Barakat           \unskip,\iYALE
% ...
}
%
% \vskip \baselineskip                   % \bigskip did not work
% \centerline{(The SLD Collaboration)}   % include collaboration name
  \vskip \baselineskip                   % \bigskip did not work
  }   % end of author list
  \address{                        % institution address list
  \baselineskip=.75\baselineskip   % shrink the interline spacing
  \iADEL
     Adelphi University,
     Garden City, New York 11530 \break
  \iBOL
     INFN Sezione di Bologna,
     I-40126 Bologna, Italy \break
  \iBU
     Boston University,
     Boston, Massachusetts 02215 \break
  \iBRUN
     Brunel University,
     Uxbridge, Middlesex UB8 3PH, United Kingdom \break
  \iCIT
     California Institute of Technology,
     Pasadena, California 91125 \break
  \iUCSB
     University of California at Santa Barbara,
     Santa Barbara, California 93106 \break
  \iUCSC
     University of California at Santa Cruz,
     Santa Cruz, California 95064 \break
  \iCIN
     University of Cincinnati,
     Cincinnati, Ohio 45221 \break
  \iCSU
     Colorado State University,
     Fort Collins, Colorado 80523 \break
  \iCOLO
     University of Colorado,
     Boulder, Colorado 80309 \break
  \iCOL
     Columbia University,
     New York, New York 10027 \break
  \iFER
     INFN Sezione di Ferrara and Universit\`a di Ferrara,
     I-44100 Ferrara, Italy \break
  \iFRA
     INFN  Lab. Nazionali di Frascati,
     I-00044 Frascati, Italy \break
  \iILL
     University of Illinois,
     Urbana, Illinois 61801 \break
  \iLBL
     Lawrence Berkeley Laboratory, University of California,
     Berkeley, California 94720 \break
  \iMIT
     Massachusetts Institute of Technology,
     Cambridge, Massachusetts 02139 \break
  \iMASS
     University of Massachusetts,
     Amherst, Massachusetts 01003 \break
  \iMISS
     University of Mississippi,
     University, Mississippi  38677 \break
  \iNAG
     Nagoya University,
     Chikusa-ku, Nagoya 464 Japan  \break
  \iOREG
     University of Oregon,
     Eugene, Oregon 97403 \break
  \iPAD
     INFN Sezione di Padova and Universit\`a di Padova,
     I-35100 Padova, Italy \break
  \iPERU
     INFN Sezione di Perugia and Universit\`a di Perugia,
     I-06100 Perugia, Italy \break
  \iPISA
     INFN Sezione di Pisa and Universit\`a di Pisa,
     I-56100 Pisa, Italy \break
  \iRUT
     Rutgers University,
     Piscataway, New Jersey 08855 \break
  \iRAL
     Rutherford Appleton Laboratory,
     Chilton, Didcot, Oxon OX11 0QX United Kingdom \break
  \iSLAC
     Stanford Linear Accelerator Center, Stanford University,
     Stanford, California 94309 \break
  \iTENN
     University of Tennessee,
     Knoxville, Tennessee 37996 \break
  \iTOH
     Tohoku University,
     Sendai 980 Japan \break
  \iVAND
     Vanderbilt University,
     Nashville, Tennessee 37235 \break
  \iWASH
     University of Washington,
     Seattle, Washington 98195 \break
  \iWISC
     University of Wisconsin,
     Madison, Wisconsin 53706 \break
  \iYALE
     Yale University,
     New Haven, Connecticut 06511 \break
  }   % end of address list

\vfill\eject

The Standard Model of electroweak
interactions is a gauge theory based on the
SU(2)$_L \times $U(1) group, which
unifies the electromagnetic and weak interactions.
Four gauge bosons
%characterize
constitute the electroweak theory:
the massless photon ($\gamma$) and three massive
bosons, the $W^+,W^-,$ and \z0.
The massive bosons acquire mass and the neutral bosons
($\gamma$ and \z0) are mixed through spontaneous symmetry
breaking.
The photon and the \z0 mixing
is described by a single
parameter, $sin^2\theta_W$.
While the electromagnetic interaction (the fermion-photon
coupling) conserves parity, the
fermion couplings to the \z0,
having both vector and axial vector components,
do not.
These components are specified as a function of
$sin^2\theta_W$ within the Standard Model.
Precision measurements of the fermion vector
and axial vector couplings to the \z0
are a stringent test of the electroweak model.
Deviations from the electroweak theory
may result from physics beyond the Standard Model.

The SLD Collaboration has recently performed the most
precise single measurement of the effective electroweak
mixing angle, $sin^2\theta_W^{\rm eff}$, by measuring the
left-right cross section asymmetry ($A_{LR}$) in $Z$ boson production
at the \z0 resonance%
\REF\alr{SLD Collaboration, K. Abe \it et al.\rm,
Phys. Rev. Lett. {\bf 73}, 25 (1994).} [\alr].
The left-right
cross section asymmetry
is a measure of the initial state
electron coupling to the \z0, which allows all visible fermion final
states to be included in the measurement.  For simplicity,
the \ep final state (Bhabha scattering) is omitted
in the $A_{LR}$ measurement due to the
dilution of the asymmetry from the large QED contribution
of the t-channel photon exchange.  In this Letter, we present
two new results:
the first measurement of the left-right
cross section asymmetry in polarized Bhabha scattering
                  ($A_{LR}^{\epo}(\act)$),
and measurements of the effective electron coupling parameters
based on a combined analysis of the
$A_{LR}$ measurement [\alr] and
the Bhabha cross section and angular distributions.
The vector coupling measurement is the most
precise yet presented.

In the Standard Model, measuring
the left-right asymmetry yields a value
for the quantity $A_e$, a measure of the degree of parity violation
in the neutral current, since:
$$
      A_{LR} = A_e =
%{{g_L^e}^2 - {g_R^e}^2\over{{g_L^e}^2 + {g_R^e}^2}} =
{2v_e a_e\over{{v_e}^2 + {a_e}^2}}=
{2[1-4sin^2\theta_W^{\rm eff}]\over{1+[1-4sin^2\theta_W^{\rm eff}]^2}},
\eqn\eqaf
$$
where the effective electroweak mixing parameter is
defined %
\REF\conven{This follows the convention used by the LEP
Collaborations in Phys. Lett. B {\bf 276}, 247 (1992).}
[\conven]
as $ sin^2\theta_W^{\rm eff} =
{1\over{4}}(1-v_e/a_e)$, and $v_e$ and $a_e$ are the effective vector and
axial vector electroweak coupling parameters of the electron.
The partial width for \z0
decaying into \ep is dependent on the coupling parameters :
$$
\Gamma_{ee} = {G_F M_Z^3\over{6\sqrt{2}\pi}}
({v_e}^2 + {a_e}^2) (1 + \delta_e),
\eqn\eqgamdef
$$
where $\delta_e
 = {3\alpha \over{4\pi}}$
is the correction for final state radiation.
$G_F$ is the Fermi coupling constant and $M_Z$ is the
$Z^0$ boson mass.
By measuring $A_e$ and $\Gamma_{ee}$, the above equations can
be utilized to extract $v_e$ and $a_e$.

The data presented in this letter were collected during the
1992 and 1993 runs of the SLAC Linear Collider (SLC), which
collides unpolarized positrons with longitudinally polarized electrons
at a center of mass energy near the \z0 resonance %
\REF\slc{N. Phinney,  Int. J. Mod. Phys. A, Proc. Suppl.
{\bf 2A}, 45 (1993)     }
[\slc].
The luminosity-weighted center of mass energy was measured to
be $ 91.55 \pm 0.02 $ GeV for the 1992 run and $91.26 \pm 0.02$
GeV  for the 1993 run %
\REF\ebeam{J. Kent \it et al.\rm, Report No. SLAC-PUB-4922,
March 1989.}
[\ebeam].
The luminosity-weighted
electron beam polarization ($<\!{\cal P}_e\!>$) was measured to be
($22.4 \pm 0.7$)\% for the 1992 run and ($63.0 \pm 1.1$)\% for the
1993 run
[\alr]%
\REF\pol{D. Calloway \it et al.\rm, Report No. SLAC-PUB-6423,
in preparation.}
[\pol].

The analysis presented here utilizes
the calorimetry systems of the SLD detector %
\REF\sld{The SLD Design Report, SLAC Report 273, 1984.}
[\slc].
Small angle coverage (28-65 mrad from the beamline) is provided
by the finely-segmented
silicon-diode/tungsten-radiator luminosity calorimeters (LUM) %
\REF\lum{S.C. Berridge \it et al.\rm, IEEE Trans. Nucl. Sci.
{\bf 39}, 1242 (1992).}
[\lum].
The LUM
           measures small angle Bhabha scattering, thereby providing
   both the absolute luminosity and
a check that
the left-right luminosity asymmetry is small.
Events at larger angles from the beamline
%($> 10^\circ$)
($> 200\,\rm mrad$) are measured with the
liquid argon calorimeter (LAC) %
\REF\lac{D. Axen \it et al.\rm, Nucl. Inst. Meth. {\bf A328}, 472
(1993).}
[\lac].
The LAC is comprised of a fine sampling electromagnetic section followed
by a coarse sampling hadronic section.  The electromagnetic section
is 21 radiation lengths in depth for normal incident particles
and contains $\sim\!99\%$
 of a 50 GeV electron
shower.  The LAC covers 98\% of the solid angle with projective
tower segmentation.

The LUM detectors surround the
beampipe on both sides of the interaction point.
Event selection criteria are designed to discriminate
high energy electromagnetic showers from background.
Selected events
are narrow
and deposit energy throughout the depth of the calorimeter while
the low energy beam backgrounds from the SLC are diffuse.
Electron position is inferred from the energy sharing between
adjacent silicon pads.

To minimize systematic uncertainties in the LUM
due to detector misalignment and
the location of the interaction point,
%to the
%% steeply falling
%Bhabha cross section which falls steeply with
%scattering angle, $\theta$,
we employ a ``gross-precise''
method %
\REF\rgp{J. Hylen \it et al.\rm, Nucl. Instr. and Meth. {\bf A317},
453 (1992) and references therein.}
[\rgp]
in the small angle measurement, which uses a larger fiducial
region on one end of the detector than the other.
The gross-precise method employs the two
luminosity monitors
as single-arm spectrometers.  Bhabha events are
identified                        in both detectors, but the
events            are counted based on the location of each
shower in the respective
detector.  In each detector, a
tight fiducial region and a loose fiducial region are defined.
The tight fiducial region is defined by silicon pad boundaries,
where the position resolution is optimal %
\REF\rbmt{S.C. Berridge \it et al.\rm, IEEE Trans. Nucl. Sci.
{\bf 37}, 1191 (1990).}
[\rbmt].
Events in which both the electron and positron showers are
within the tight fiducial region are labeled as ``precise''
Bhabhas and counted with weight 1.  Events in which one of the
two showers is inside the tight fiducial region and the
other shower is outside the tight fiducial
region  are labeled as ``gross'' events and given
weight $1/2$.
With the gross-precise method applied to these data,
the misalignment error on the effective number of calculated events
is negligible %
\REF\thesis{K.T. Pitts, Ph.D. Thesis, University of Oregon,
SLAC Report 446 (1994).}
[\thesis].
% would be
%An effective number of events can then
%be calculated
%$ n_{\rm eff} = n_{precise} + {n_{gross}\over{2}} $.
The effective cross section is calculated by using the
Monte Carlo programs BABAMC %
\REF\bmc{F.A. Berends, R. Kleiss, and W. Hollik,
Nucl. Phys. {\bf B304},
712 (1988).}
[\bmc]
and BHLUMI %
\REF\mc{S. Jadach and B.F.L. Ward,
                Phys.  Rev {\bf D40}, 3582 (1989);
S. Jadach \it et al.\rm, Phys. Lett. B {\bf 268}, 253 (1991).}
[\mc].
Detector simulation is performed with GEANT %
\REF\geant{R. Brun \it et al.\rm, CERN--DD/78/2 (1978).}
[\geant]
and
the electromagnetic showers are parameterized using the
GFLASH algorithm %
\REF\gflash{G. Grindhammer \it et al.\rm, Nucl. Inst. Meth. {\bf
A290},469(1990).}
[\gflash].

The overall errors for the physics measurements to be presented are
limited by small statistics.  For this reason, our systematic
error analysis of the luminosity measurement is  conservative.
A detailed description of the systematic error analysis
for the luminosity measurement is given elsewhere
   [\thesis].
The total systematic uncertainty is             0.93\%, which
is
composed
of 0.88\% experimental and 0.3\% theoretical
uncertainty.  The experimental systematic
error is limited by
the size of the data set.
The integrated luminosity
is ${\cal L} = 385.4  \pm 2.5 \ \rm (stat)
\pm 3.6 \  (sys)\ nb^{-1}$ for the 1992 polarized SLC run and
 ${\cal L} = 1781.1 \pm 5.1\ \rm (stat) \pm 16.6\  (sys)
\ nb^{-1}$ for the 1993 SLC run.

The wide angle Bhabha selection algorithm makes
use of the distinct topology of the \ep final state.
%Events are required to have two large clusters of electromagnetic
%energy which are nearly back-to-back in the detector.  These clusters
%are required to have deposited very little energy in the hadronic
%calorimeter, as well as very little energy beyond the two primary clusters.
%For each event candidate, energy clusters are reconstructed in the
%LAC.
Selected events are required to possess two clusters which contain
at least 70\% of the center of mass energy
%(this cut is relaxed in the endcap region)
and manifest a normalized energy
imbalance of less than 0.6 %
\REF\imb{The energy imbalance is defined as the normalized vector
sum of the energy clusters as follows, $E_{imb} =
|\sum{\vec{E}_{cluster}}|/\sum{|\vec{E}_{cluster}|}$.}
[\imb].
The two largest energy clusters
are also required to have less than 3.8 GeV of energy in the hadronic
calorimeter.  The total number of reconstructed clusters found in the
event must be less than 9.   Collinearity in  the final state
is controlled by requiring the
absolute value of the rapidity sum of the two main clusters to
be less than 0.30.

The efficiency and contamination for the wide angle events
are calculated from Monte Carlo
simulations.
Corrections are applied as a function of scattering angle to account
for angle-dependent changes in response.  The \bha\ process at
large angles is simulated with BHAGEN %
\REF\bhagen{M. Caffo, H. Czy\.z and E. Remiddi,
Nuovo Cim. {\bf 105A}, 277 (1992).}
[\bhagen].
\REF\radcor{F.A. Berends and R. Kleiss,
Nucl.Phys. {\bf B186}, 22 (1981).}
\REF\koralz{S. Jadach, B.F.L. Ward and Z. Wa\c s,
Comp. Phys. Commun. {\bf 66}, 276 (1991).}
The efficiency for accepting wide angle
         \bha \ events is found to be
86.7\% overall and 93\%
in the central region of the detector,
                    with the largest inefficiency arising
from events which
enter the gaps between adjacent liquid argon modules.  The efficiency
falls off in the forward regions due to materials in front of the
calorimeter from the interior detector systems.

Two small
sources of contamination are $e^+e^-\rightarrow \gamma \gamma$
(1.25\%) and
$e^+e^- \rightarrow \tau^+\tau^-$ (0.28\%).
The Monte Carlo programs RADCOR and
KORALZ are used to calculate these contributions [\radcor,\koralz].
              Other sources of contamination such as hadronic decays
of the \z0, two-photon, cosmic rays and beam background were all found
to give negligible contributions.

Tables~I and II show the number of events
accepted, by beam helicity, for the 1992 and 1993 SLC runs.
The raw asymmetry is defined as:\break
\centerline{\alrraw  $ = <\!\Pe\!>$ \alrbb
$ = (N_L - N_R)/(N_L + N_R)$,}
where
$N_L$($N_R$) is the number of events tagged with a
left-(right-) handed electron beam as a function of the $|cos\theta|$,
where $\theta$ is the
center-of-mass scattering angle for the \ep system after initial state
radiation.
Aside from the charge ambiguity which is unresolved by the
calorimeter measurement,
the center-of-mass scattering angle is derived trivially from the
measured electron and positron laboratory scattering angles.
The angular
regions in the table are chosen to emphasize the different
regimes of the \bha\ distribution: for    $ |cos\theta| < 0.7 $
the s-channel \z0 decay dominates; from 0.7 to 0.94 the s-channel
\z0 decay, the t-channel photon exchange and the interference
between those two interactions all contribute; for
$|cos\theta| > 0.94$, the t-channel photon exchange dominates.
The region of $0.998<|cos\theta|<0.9996$ is that which is covered
by the LUM.
The expected     asymmetry (\alrbb) is largest
at $cos\theta = 0$,
and may be approximately written as
\alrbb$ = A_e(1-\ft)$, where $\ft$ represents the
t-channel contribution.  For the region $ |cos\theta| < 0.7$,
$< f_t > \simeq 0.12$.
The expected asymmetry
falls to very small values ($\sim\! 10^{-4}$) in the small angle
region where the t-channel photon exchange dominates.

To extract $\Gamma_{ee}$ and $A_e$, the data are fit
to the differential \ep cross section
using the maximum likelihood method.
Two programs are used to
calculate the differential \ep cross section: EXPOSTAR %
\REF\expo{D. Levinthal, F. Bird, R.G. Stuart and B.W. Lynn, Z. Phys. C
{\bf 53}, 617 (1992).}
[\expo]
and, as a cross check, DMIBA %
\REF\dmiba{P. Comas and M. Martinez, Z. Phys. C {\bf 58}, 15 (1993).}
[\dmiba].
The EXPOSTAR program calculates the differential cross sections
within the framework of the Standard Model.  The DMIBA program
calculates the differential \ep \ cross section in a model independent
manner.
To extract the maximal
amount of information from the differential polarized
Bhabha scattering distribution, the fit is performed over the
entire angular region accepted by the LAC ($|cos\theta| < 0.98$).
No t-channel subtraction is performed. All ten lowest order terms in the
cross section are included in the fit:  the four pure s-channel and
t-channel terms for
photon and \z0 exchange, and the six interference terms %
\REF\terms{see e.g. M. Greco, Nucl. Phys. {\bf B177}, 97 (1986).}
[\terms].
The fit also includes initial state radiation.  Since the
measurement is calorimetric it is insensitive to final state radiation.

The partial width
$\Gamma_{ee}$ is extracted from the data in two ways: (1) using the
full fit to the differential cross section for $|cos\ \theta| \le $ 0.98,
and (2) measuring the cross section in
the central region ($|cos\ \theta| < 0.6$)
where the systematic errors are smaller,    yielding a more precise
measurement.  For the fits we use $M_Z  = 91.187 \ \rm GeV/c^2$ and
$\Gamma_Z = 2.489 \ \rm GeV/c^2$ %
\REF\lepave{The LEP Collaborations and The LEP Electroweak Working
Group, Report No. CERN-PPE-93-157, August, 1993.}
[\lepave].
Figure 1 shows
the fit to the full  \bha\  distribution.
This fit has a $\chi^2 = 51.6$ for 39 degrees of freedom, yielding
$\Gamma_{ee} = 83.14 \pm 1.03\ \rm (stat) \pm 1.95\  (sys)\ \rm{MeV}$.
The 2.4\% systematic error is dominated (2.1\%) by the
uncertainty in the efficiency correction factors
in the angular region 0.6 $<|cos\theta|< $ 0.98, where
the LAC response is difficult to model due to materials
from interior detector elements%
 [\thesis].

A more precise determination of
$\Gamma_{ee}$ was performed  using only the central
region of the LAC ($|cos\theta|<0.6$)
 and the small angle region in the LUM %
\REF\johny{see similar analysis in
J.M. Yamartino, Ph.D. Thesis, MIT,
SLAC Report 426, February 1994.}
[\johny].
The program MIBA %
\REF\miba{M. Martinez and R. Miquel, Z. Phys. C {\bf 53}, 115 (1992).}
[\miba]
is then
used to calculate \gee \ based on the total measured cross section
within the defined fiducial region.  From this method, we find:
$$
\Gamma_{ee} = 82.89 \pm 1.20\ \rm (stat) \pm 0.89\  (sys)\ \rm{MeV}.
$$
The loss in statistical precision of the limited fiducial region
is more than compensated by
the improvement in the systematic error.  The 1.1\% systematic
error is dominated by the accuracy of
the detector simulation (0.74\%) and the
uncertainty in the absolute luminosity (0.52\%).
Other contributions are the uncertainty in the contamination (0.3\%),
the uncertainty in $M_Z$ and
$\Gamma_Z$ (0.3\%), the accuracy of the cross section
calculation (0.3\%), and the center of mass energy uncertainty (0.2\%),
%The following factors contribute to the systematic
%uncertainty:  the correction factors, the luminosity measurement,
%the center of mass energy and beam energy spread, $M_Z$ and
%$\Gamma_Z$, and the accuracy of the calculation.

To extract $A_e$ from the Bhabha events,
the right- and left-handed
              differential \bha\ cross sections are
fit directly for $v_e$ and $a_e$
using EXPOSTAR.  This yields
$$
A_e = 0.202 \pm 0.038\ \rm (stat) \pm 0.008 \ (sys).
$$
Figure 2 shows the measured left-right cross section
asymmetry for \bha\
                  ($A_{LR}^{\epo}(\act)$)
compared to the fit.
The fit shown in Figure 2
has a  $\chi^2$  of 4.36 for 5 degrees of freedom.
The measurement of $A_e$ is limited by the
statistical uncertainty.
The 3.8\% systematic
is dominated by a 3.2\% uncertainty in the
angle-dependent response correction factors.
The polarization uncertainty contributes 1.7\% and asymmetry
factors from the SLC contribute 0.06\% as discussed in
Refs.~[\alr] and [\thesis].  Other systematic error
contributions to $A_e$ are the beam energy spread and center
of mass energy uncertainty (0.25\%), the accuracy of the
EXPOSTAR program (0.7\%) and the uncertainty on the
\z0 mass and width (0.7\%).
%error has contributions from the factors listed above for $\Gamma_{ee}$,
%as well as the uncertainty in the electron beam polarization measurement
%and the left-right asymmetry factors discussed in Refs.~[\alr] and
%[\thesis].

The results for \gee \ and $A_e$ from above
may now be used in equations~\eqaf\  and \eqgamdef \
to extract the effective
vector and axial vector couplings to the \z0:
$
v_e = -0.0507 \pm 0.0096\ \rm  (stat) \pm 0.0020\  (sys),
\it a_e \rm = -0.4968 \pm 0.0039\ \rm (stat) \pm 0.0027\  (sys),
$
where lower energy \ep annihilation data have been utilized to
assign $|v_e| < |a_e| $,
and $\nu_e e$ scattering data have been utilized to
establish $v_e <0$ and $a_e<0$ \REF\vasign{
S.L. Wu, Phys. Rep. \bf 107\rm, 59 (1984).}
[\vasign].
Figure 3   shows the one standard deviation (68\%)
contour
for these electron vector and axial vector coupling measurements.
Most
of the sensitivity to the electron vector coupling
    and, hence, $sin^2\theta_W^{\rm eff}$ arises from the
measurement of $A_e$, while the sensitivity to the
axial vector coupling            arises from $\Gamma_{ee}$.
Also shown  are standard model calculations using
the program ZFITTER %
\REF\zfitter{D. Bardin \it et al.\rm,
Report No. CERN-TH-6443-92, May 1992.}
[\zfitter].

The effective electroweak mixing angle represented by
these vector and axial vector couplings is:
$$
sin^2\theta_W^{\rm eff} = 0.2245 \pm
0.0049\ \rm (stat) \pm 0.0010\  (sys).
$$
\noindent
We reiterate that this measurement derives strictly from the Bhabha
events.

The SLD Collaboration has published a more precise measurement of
$A_e$ from the left-right cross section asymmetry ($A_{LR}$)
measurement %
[\alr].
Combining the Bhabha results with the SLD measurement of $A_{LR}$
gives:
$$
v_e = -0.0414 \pm 0.0020  \hskip .6 in
a_e = -0.4977 \pm 0.0045,
$$
the most precise          measurement of the electron
vector coupling to the \z0 published to date.
The $v_e$, $a_e$ contour including the $A_{LR}$ measurement
is also shown in Figure 3,   demonstrating the increased
sensitivity
in $v_e$ from $A_{LR}$.
%Figure~\ref{fig:1sig} also shows that the measurement of $A_e$
%from polarized \bha\ agrees with the measured value of \alr.
%Figure~\ref{fig:vecomp} shows the result presented
%here for the electron vector coupling to the \z0 compared
%with other measurements.   The second entry in Figure~\ref{fig:vecomp}
%is the SLD measurement of $A_{LR}$ combined with the
%measurement of $\Gamma_{ee}$ presented here.  The LEP results
%for each experiment come from either the electron forward-backward
%asymmetry ($A_{FB}$) or the forward-backward asymmetry in tau
%decays combined with $\Gamma_{ee}$.
%The value of $A_e$ measured from \bha\ agrees with the
%SLD measurement of the left-right cross section asymmetry [\alr].
The LEP average
for the electron coupling parameters to the \z0 are
$v_e = -0.0370 \pm 0.0021$
and $a_e = -0.50093 \pm 0.00064$ %
\REF\lepgla{D. Schaile, 27th International Conference on High Energy
Physics, Glasgow, Scotland, July, 1994.}
[\lepgla].

In summary,  the effective electron coupling parameters have
been determined with a new method which
combines  the left-right cross section asymmetry ($A_{LR}$)
with the  polarized Bhabha scattering differential cross section.
The effective electron vector coupling to the \z0 is determined
with the    best precision to date.

We thank the personnel of the SLAC accelerator department and the
technical staffs of our collaborating institutions for their outstanding
efforts on our behalf.

\bigskip
%\vfill\eject
\refout
\vfill\eject

\noindent{\bf Table I.} Number of accepted Bhabha
events and their raw asymmetry for the 1992 run.  The average
electron beam polarization was 22.4\%.
\bigskip
\begintable
$\vert cos\theta\vert$| left-handed | right-handed |
\alrraw  $ = <\!\Pe\!>$ \alrbb    \crthick
$                   < 0.70$ | 157 | 137 | 0.068 $\pm$ 0.058 \cr
$0.70          -      0.94$ | 208 | 205 | 0.0073 $\pm$ 0.049 \cr
$0.94          -      0.98$ | 305 | 318 | $-$0.021 $\pm$ 0.040 \cr
$0.998         -       0.9994$ | 12,395 | 12,353 | 0.0017 $\pm$ 0.0064
\endtable

\vskip 1truein
\noindent{\bf Table II.} Number of accepted Bhabha
events and their raw asymmetry for the 1993 run.  The average
electron beam polarization was 63.0\%.
\bigskip
\begintable
$\vert cos\theta\vert$ | left-handed | right-handed |
\alrraw  $ = <\!\Pe\!>$ \alrbb    \crthick
$              < 0.70$ |  864 | 702 | 0.103 $\pm$ 0.0253 \cr
$0.70       -         0.94$ | 1,039 | 946 | 0.047  $\pm$ 0.022 \cr
$0.94       -         0.98$ | 1,566 | 1,479 | 0.029 $\pm$ 0.018 \cr
$0.998      -          0.9996$ | 93,727 | 94,319 | $-$0.0032 $\pm$ 0.0023
\endtable
\vfill\eject
\noindent{\bf Figure 1.}  Differential angular distribution
for \bha.   The points are the corrected data, the dashed
line is the fit.
\vskip 1truein
\noindent{\bf Figure 2.}  Left-right asymmetry, \alrbb, for polarized
\bha.  The points are the corrected data, the dashed curve is the fit.
\vskip 1truein
\noindent{\bf Figure 3.}  One-sigma (68\%) contour in the
$a_e$,$v_e$ plane.  The large ellipse is for \bha, the smaller
ellipse includes the measurement of $A_{LR}$.  The shaded region
represents the Standard Model calculation for
130 GeV $< m_{top}  <$ 250 GeV and
50 GeV $<  M_{Higgs} <  1000 $ GeV.
\vfill\eject
% \figout    %
% \vfil\end  %
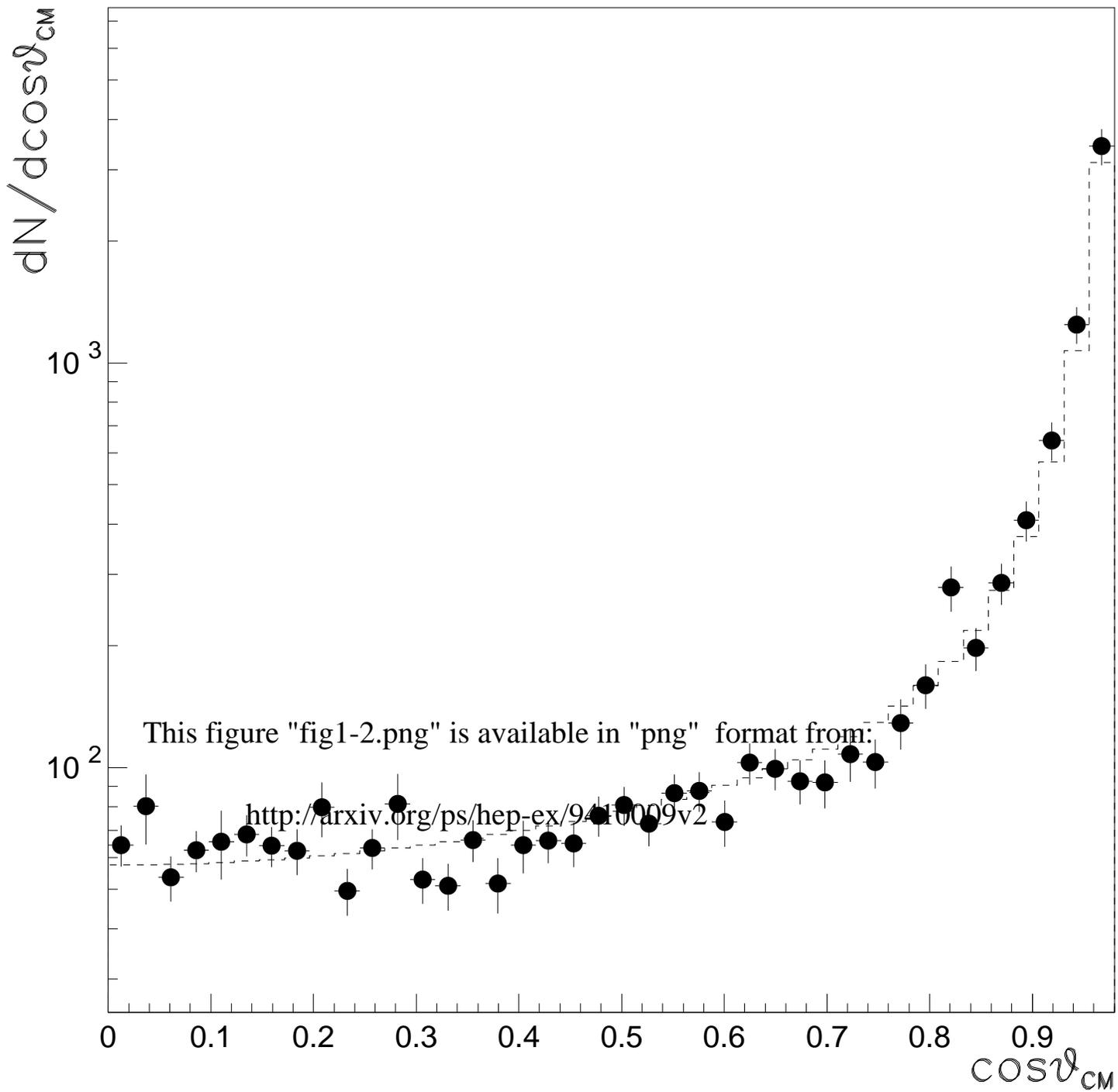
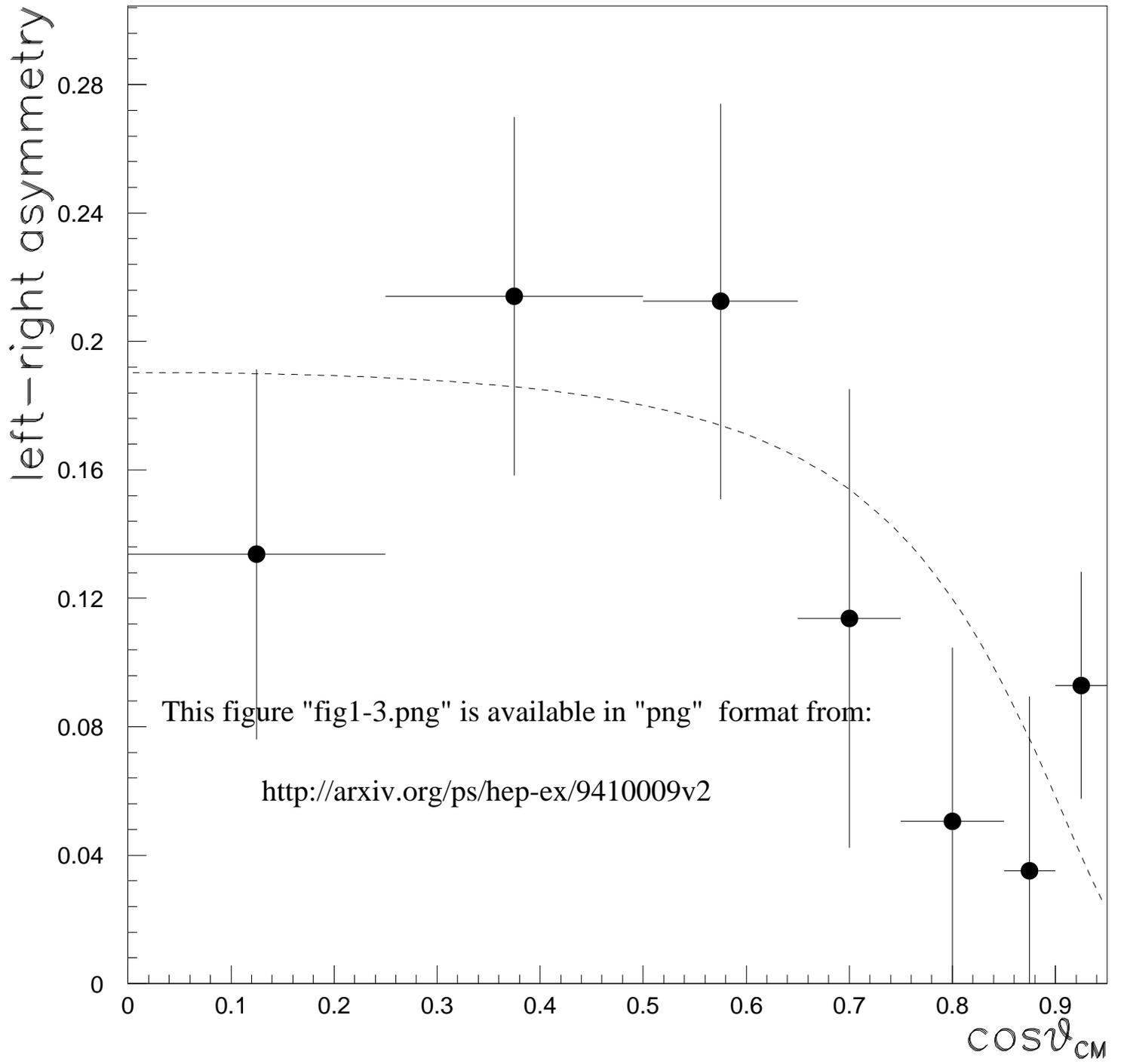
\bye